\documentclass{jpp}
\usepackage{graphicx}% Include figure files
\usepackage{dcolumn}% Align table columns on decimal point
\usepackage{bm}% bold math
\usepackage{hyperref}% add hypertext capabilities
\usepackage{epstopdf}
\usepackage{amsmath}
\graphicspath{{./fig/}}

\newcommand{\EQ}{\begin{equation}}
\newcommand{\EN}{\end{equation}}
\newcommand{\BB}{\bm{B}}
\def\cs{c_{\rm s}}

\newcommand{\AAA}{\bm{A}}
\newcommand{\jj}{\mbox{\boldmath $j$} {}}

\newcommand{\dd}{{\rm d} {}}

\newcommand{\vv}{\mbox{\boldmath $v$} {}}

\newcommand{\pder}[2]{\frac{\partial#1}{\partial#2}}

\newcommand{\refbold}[1]{{#1}}

\title{Resistive evolution of toroidal field distributions\\ and their relation to magnetic clouds}% Force line breaks with \\
\date{\today}% It is always \today, today,

\author{C. B. Smiet\aff{1}\aff{2}\corresp{\email{csmiet@pppl.gov}}
H. J. de Blank\aff{3}
T. A. de Jong\aff{2}
D. N. L. Kok\aff{2}
D. Bouwmeester\aff{2}\aff{4}
}
\affiliation{\aff{1}Princeton Plasma Physics Laboratory, Princeton University, Princeton, New Jersey, USA
\aff{2}Huygens-Kamerlingh Onnes Laboratory, Leiden University, P.O.\ Box 9504, 2300 RA Leiden, The Netherlands
\aff{3}DIFFER - Dutch Institute for Fundamental Energy Research, De Zaale 20, 5612 AJ Eindhoven, the Netherlands
\aff{4}Department of Physics, University of California Santa Barbara, Santa Barbara, California, 93106, USA}

\begin{document}
\maketitle

\begin{abstract}
    We study the resistive evolution of a localized self-organizing magnetohydrodynamic
    equilibrium.
    In this configuration the magnetic forces are balanced by a pressure force caused
    by a toroidal depression in the pressure. Equilibrium is attained when this low pressure
    region prevents further expansion into the higher-pressure external plasma.
    We find that, for the parameters investigated, the resistive evolution of the structures
    follows a universal pattern when rescaled to resistive time.
    The finite resistivity causes both a decrease in the magnetic field strength and a finite slip
    of the plasma fluid against the static equilibrium.
    This slip is caused by a Pfirsch-Schl\"uter type diffusion, similar to what is seen in
    tokamak equilibria.
    The net effect is that the configuration remains in Magnetostatic equilibrium whilst it
    slowly grows in size.
    The rotational transform of the structure becomes nearly constant throughout the entire
    structure, and decreases according to a power law.
    \refbold{In simulations} this equilibrium is observed when highly tangled field lines
    relax in a
    high-\refbold{pressure (relative to the magnetic field strength)} environment, a situation that occurs when the twisted field of a coronal loop is
    ejected into the interplanetary solar wind.
    In this paper we relate this localized MHD equilibrium to magnetic clouds in the solar
    wind.
\end{abstract}

\section{Introduction}
Spontaneous self-organization of magnetized plasma lies at the basis of many fascinating
phenomena in both fusion reactor operation and astrophysical plasma observations.
In such situations, magnetic helicity in the plasma is of crucial
importance in determining the evolution of the system.
Magnetic helicity is an integral quantity calculated by $H_{\rm m}= \int \AAA\cdot\BB
\mathrm{d}^3x$ where $\AAA$ is the vector potential and $\BB=\nabla\times\AAA$ is the magnetic
field.
Helicity was given its name by~\citet{moffatt1969degree} who recognized its
topological interpretation;  a
measure of the self- and interlinking of magnetic field lines. This notion was extended to
non-closing and ergodic field lines by~\citet{arnold1986asymptotic}. In a perfectly conducting
plasma, the magnetic field can be seen as being `frozen in' and is advected with the fluid
motion\citep{alfven1943existence,BatchelorFrozeIn1950RSPSA}. Since the fluid motions can then only distort and reshape the
magnetic field lines, but cannot break or cause unlinking, the conservation of helicity is
easily understood from a topological perspective.

In perfectly conducting plasma helicity is
exactly conserved~\citep{berger1984topological}, whilst in resistive plasma the rate of energy
decay is strongly constrained by its presence~\citep{del2010magnetic}.
The most famous example where helicity determines a self-organizing process is the Taylor
conjecture~\citep{Taylor1974,taylor1986relaxation} which
states that the magnetic field in a toroidally bounded plasma relaxes to a linear-force free
state (shown by \citet{woltjer1958theorem} to be the lowest-energy state) with
exactly the same helicity as it started with.

But what state is achieved when a helical plasma relaxes in an environment without a boundary?
If the plasma's fluid pressure is high compared to the magnetic pressure (high plasma
$\beta=\frac{p}{B^2/2}$), the linking in the initial field gives rise to a self-organized MHD
equilibrium where field lines lie on nested toroidal magnetic surfaces~\citep{smiet2015self}.
The approximately axisymmetric field is in a Grad-Shafranov
equilibrium~\citep{shafranov1966plasma}, and the
Lorentz force is balanced by a gradient in pressure.
The pressure is lowest on the magnetic axis, which is the field line lying at the center of
the nested toroidal magnetic surfaces.
In this structure the rotational transform is nearly
constant from surface to surface.
As a consequence the magnetic field line structure is
topologically similar to the mathematical
structure of the Hopf fibration~\citep{Hopf1931} or its generalization to torus
knots~\citep{arrayas2014class, smiet2015self}.
It should be noted that the Hopf structure has previously also been used
in a beautiful paper by
Finkelstein and Weil~\citep{finkelstein1978magnetohydrodynamic} to generate linked and knotted
magnetic fields for astrophysics.
The relaxation of the Hopf field to the Grad-Shafranov equilibrium has been shown using
topology conserving relaxation in our recent paper~\citep{smiet2016ideal}.
It is remarkable that this structure is
obtained for a wide class of initially helical fields; it emerges from
trefoils~\citep{smiet2015self}, twisted rings, and even Borromean
linked flux tubes~\citep{Candelaresi2011helical}.

The robust generation of this ordered magnetic structure makes it natural to assume that
a similar process emerges after the twisted magnetic field of a coronal loop is ejected
into the high-beta interplanetary plasma of the solar wind. \refbold{The pressure in the solar wind at 1 A.U. is of the order of $p=1.4\times
10^{-11} N m^{-2}$ and the field strength  $B=6\times 10^{-9} T$ such that the plasma
$\beta=2\mu_0 p/B^2 \simeq 1$~\citep{goedbloed2004principles}.}
Such events are called Coronal
Mass Ejections (CME's), and CME's are correlated with the observation of magnetic
clouds~\citep{burlaga1991magnetic}.
A magnetic cloud is a localized magnetic structure in the interplanetary plasma with increased
magnetic field strength, and where the direction of the field varies by a large
angle~\citep{burlaga1991magnetic}.
These magnetic signatures are observed by interplanetary sattelites with increasingly accurate
magnetic instruments~\citep{raghav2018first}.
Unfortunately due to the low
density of probes in the interplanetary medium, high-resolution measurement of the complete magnetic
structure in these clouds is challenging.

There are several models for the magnetic structure of these clouds published in
literature~\citep{burlaga1991magnetic}. Some models assume the field in the cloud is still
magnetically connected to the surface of the sun, but there are several models that
assume a localized magnetic field which is created by internal currents and balanced
by the external
plasma pressure~\citep{vandas1992magnetic, kumar1996interplanetary, burlaga1991magnetic, ivanov1985interplanetary,
garren1994lorentz}.
In this paper we posit the self-organized state identified in~\citep{smiet2015self} as a new
model.
This model gives different predictions for the structure from the models above.

In this paper, we study the evolution of the self-organizing equilibrium starting from a twisted flux tube.
We vary both the resistivity of the simulation as well as the amount of twist in the initial
flux tube.
The evolution of the structure is governed by two processes.
First, the lowering of the magnetic field
strength changes the equilibrium condition, such that the depression in pressure becomes smaller.
Second, finite resistance breaks the frozen-in condition, allowing the plasma
fluid to slip perpendicular to the magnetic field lines of the configuration.
The net effect of this is a fluid flow directed towards the region of lowest pressure.
The combined effect of these two processes is a structure that grows on a resistive
timescale.
The magnetic topology, characterized by the rotational transform
of the magnetic surfaces, quickly reaches a nearly flat profile with a slight positive
curvature.
The rotational transform decays according to a power law, the
characteristic exponent of which depends on the aspect ratio of the structure.

\section{Initial field}

\begin{figure}
    \begin{center}
        \includegraphics[width=.7\columnwidth]{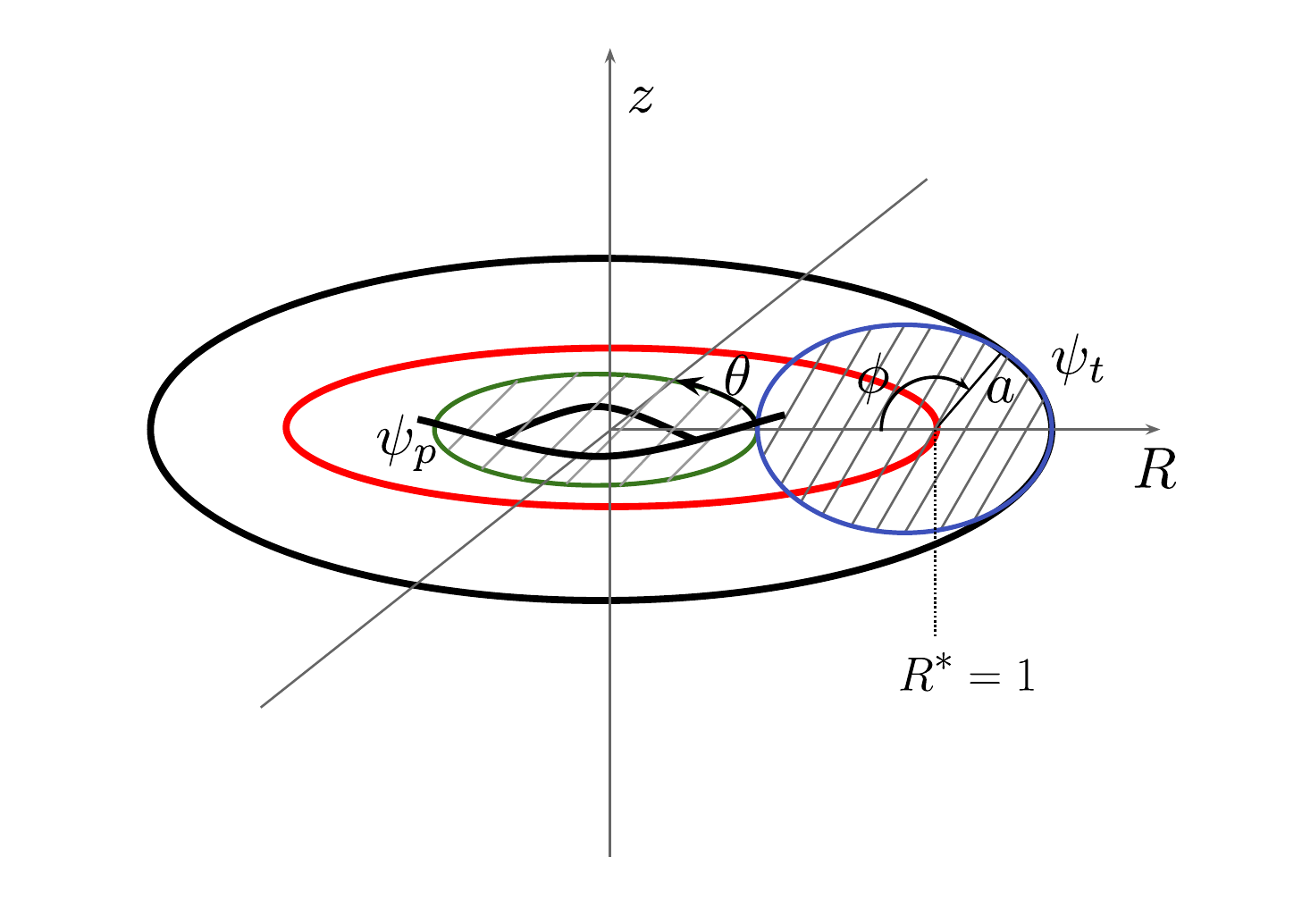}
    \end{center}
    \caption{Coordinate system used for the construction of the initial magnetic field. The
    surfaces through which the
    toroidal flux $\psi_t$ and poloidal flux $\psi_p$ are defined are shown by the blue and green
    circles respectively. $\phi$ is the coordinate
    pointing in the poloidal direction of the torus, and $\theta$ is the coordinate pointing in
    the toroidal direction. The magnetic axis is given %'initial magnetic axis' to emphasise evolution?
    by the red circle located at $R^*=1$. }
    \label{fig:coords}
\end{figure}

In our previous paper~\citep{smiet2015self}, we showed how a self-organizing equilibrium is generated
through the chaotic reconnection of linked magnetic flux rings.
The violent reconnection and high spatial gradients generated when the flux tubes
meet, necessitated a high value of resistivity and low magnetic field strength
to numerically resolve.

In this paper we use an initial condition which re-organizes in a much more ordered
fashion.
The initial field consists of a twisted flux tube with field lines lying on nested
toroidal flux surfaces with varying rotational transform.
We use cylindrical coordinates $R,z,\theta$
and flux functions $\psi_p(R,z)$ and $I(\psi_p)$.
The coordinate system is described in figure~\ref{fig:coords}.
Physically $\psi_p(R,z)$ represents the poloidal magnetic flux, passing through a circular
surface of radius $R$ around the $z$-axis (green circle).
$I$ physically represents the total poloidal current, the current through the green,
shaded circle.
The magnetic field is calculated from the flux functions by the standard methods for
axisymmetric fields~\citep{goedbloed2010advanced}:
\EQ\label{eq:field}
B_R = -\frac{1}{R} \frac{\partial \psi_p}{\partial z}, \quad B_z= \frac{1}{R} \frac{\partial
\psi_p}{\partial R}, \quad
B_\theta = \frac{1}{R} I.
\EN
Using this construction guarantees that the magnetic field is divergence free, and that field lines
lie on magnetic surfaces of constant $\psi_p$.

We choose the following flux function,
\EQ\label{eq:psi}
\psi_p =
  \begin{cases}
      B_0\cos^4\left(\frac{\pi}{2} \frac{a}{a_0}\right), & a\leq a_0\\
      0, & a >a_0
  \end{cases}
\EN
where $0<a_0<1$ and with $B_0$ a scaling parameter that sets the magnetic field strength and
\EQ
a= \sqrt{(1-R)^2+z^2},
\EN
denoting the distance from the unit circle (the circle $R=1, z=0$, indicated in red in figure
\ref{fig:coords}), which is the magnetic axis of this initial condition.

With this choice we can see that $\psi_p$ is constant and zero for $a\geq a_0$ such that, according
to equation \eqref{eq:field}, the poloidal field vanishes.
The magnetic surfaces in the region where $a < a_0$ form concentric tori with circular
cross section that enclose the magnetic axis.

We can choose any function of $\psi_p$ for the toroidal current function $I$, which here
we define as:
\EQ
I=\frac{\psi_p \pi^2 }{\imath_0^* a_0^2},
\EN
\refbold{where a scaling parameter $\imath_0^*$ is introduced, named such because
it sets the rotational transform on axis}, as we will show in Section
\ref{sec:safety}.
With this choice for $I$ the toroidal magnetic field also vanishes for $a\geq a_0$, and thus the field
describes an axisymmetric flux tube with major radius $1$ and minor radius $a_0$ with $\BB=0$ outside
of the tube. \emph{Note:} We use the convention that a subscript zero denotes a value at time
$t=0$, and a superscript asterisk denotes that the quantity is measured on the magnetic axis.

\subsection{Rotational transform profile}\label{sec:safety}
The winding of field lines in a toroidal magnetic structure is quantified by the rotational
transform $\imath$ or its inverse, the safety factor $q$.
The rotational transform geometrically represents the ratio of the number of times a field line
wraps around the poloidal direction of a torus to the number of times it winds around
the toroidal direction.
The safety factor can be calculated using the well-known formula~\citep{wesson2011tokamaks}:
\EQ\label{eq:qprof}
q=\frac{1}{2\pi} \oint \frac{1}{R}\frac{B_\theta}{B_p}\dd l,
\EN
where $B_p$ is the magnitude of the poloidal magnetic field $B_R \hat{R}+ B_z \hat{z}$, and the
integration is carried
out over a constant $\theta$ cross section of a magnetic surface $(\psi_p = \text{
    const.})$.
This integration path is indicated by the blue circle in figure \ref{fig:coords}.

\refbold{The poloidal magnetic flux enclosed between concentric magnetic
    surfaces of radii $a$ and $a$ + $\mathrm{d}a$ is $\mathrm{d}a 2 \pi R B_p.$ Hence
    conservation of poloidal flux implies that $RB_p$ is constant on each surface.
}
Evaluating the poloidal field at $z=0,
R\geq 1$, where $\pder{\psi_p}{z}=0$ and $\pder{\psi_p}{R}=\pder{\psi_p}{a}$ its value in the
direction of the poloidal vector $\phi$ is equal to
\EQ
R B_p =\pder{\psi_p}{a} =B_0 \frac{2\pi}{a_0}
\cos^3\left(\frac{a\pi}{2a_0}\right) \sin\left(\frac{a\pi}{2a_0}\right).
\EN
The \refbold{toroidal} magnetic field is:
\EQ
B_\theta = B_0 \frac{\pi^2 \cos^4\left(\frac{\pi a}{2a_0}\right)}{R \imath_0^* a_0^2}.
\EN
On the magnetic surfaces the parameter $a$ is a constant, so filling this in equation~\eqref{eq:qprof} becomes:
\EQ
q(a)=\frac{\cos\left(\frac{a\pi}{2a_0}\right)}{4\imath_0^* a_0 \sin\left(\frac{a\pi}{2a_0}\right)} \ \
\oint \frac{1}{R} \dd l.
\EN

Using $\phi$ to parametrize the integral over the surface at $a$, and the identities
$R=1+a\cos(\phi)$ and $\dd l = a\dd\phi$, we get:
\EQ
\oint \frac{1}{R} \dd l=\int_0^{2\pi} \frac{a}{1+a\cos(\phi)}\dd \phi = \frac{2\pi a}{\sqrt{1-a^2}}.
\EN
This gives us the safety factor
\EQ
q(a)=  \frac{\frac{a \pi}{2 a_0} \cot\left(\frac{a\pi}{2a_0}\right)}{\imath_0^* \sqrt{1-a^2}},
\EN
and hence a rotational transform of:
\EQ\label{eq:rottrans}
\imath(a) =\imath_0^*  \sqrt{1-a^2} \frac{\tan\left(\frac{a\pi}{2a_0}\right)}{\frac{a\pi}{2a_0}}.
\EN

The rotational transform profile is flat near the magnetic axis, and increases to infinity
when $a\rightarrow a_0$.
At the magnetic axis the rotational transform is given by
\EQ
\lim_{a\rightarrow 0} \imath(a)=\imath_0^*.
\EN

The initial condition is thus an axisymmetric, twisted magnetic flux tube lying in the
$R,\!\theta$-plane.
We can change the twist of the magnetic field lines in the initial condition by
tuning the parameter $\imath_0^*$, and the
toroidal magnetic field strength with the parameter $B_0$.

\section{Time evolution}\label{sec:observations}
\begin{figure*}
    \includegraphics[width=\columnwidth]{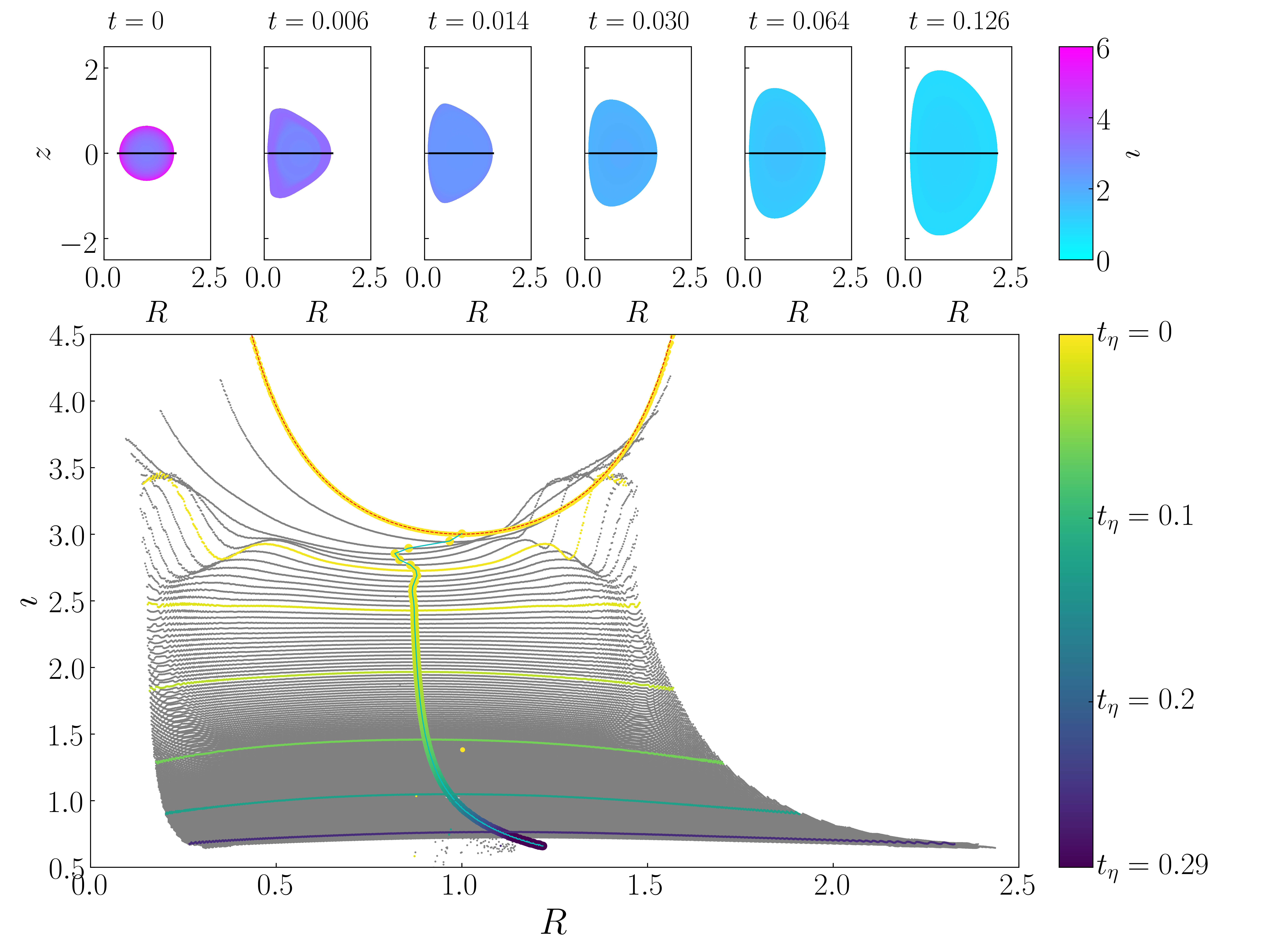}
    \caption{Resistive evolution of the magnetic structure.
    Parameters are: $\imath_0^*=3$, $B_0=0.05$ and $\eta=2\times
    10^{-4}$. Top row: Cross sections in
    the $R,\!z$-plane at different times. The color indicates the rotational
    transform of a field line starting at that position. The configuration
    is seen to first contract onto the
    $z$-axis and then slowly expand. The horizontal lines indicates the location
    where the rotational transform, shown in the bottom panel, is taken. The cross
    sections shown are at times which correspond with
    the top six colored lines in the bottom panel and are exponentially spaced.  Bottom panel:
    Evolution of the rotational transform profile and the location of the magnetic
    axis. The black line along the center is the location of the magnetic
    axis. The colored part around the black line indicates the time at which the magnetic axis
    was at that location.
    The top six colored lines correspond to the six cross sections shown in the top
    row, and their color again indicates the time at which that rotational transform profile
    was calculated. The red dashed line is the analytical result of
    equation~\eqref{eq:rottrans}. We can see that
    the configuration quickly reaches a nearly flat rotational transform profile and
    subsequently evolves self-similarly.
    }\label{fig:cudanalysis}
\end{figure*}

We simulate the time evolution of the helical magnetic fields numerically using the resistive,
viscous, compressible isothermal MHD equations.
The equations are solved using the PENCIL-CODE~(http://pencil-code.nordita.org/). This
is a highly used solver of the MHD equations on a fixed Eulerian grid often used for
astrophysical
applications~\citep{brandenburg2002hydromagnetic,haugen2004simulations,johansen2007rapid}.

The PENCIL-CODE solves the MHD equations in terms of the vector potential $\AAA$, ensuring that
the magnetic field remains divergence free.
The equation of motion solved is:
\begin{equation}\label{eqn:pencil_motion}
\frac{D \vv}{D t} = -\cs^2 \nabla \ln \rho + \jj \times \BB/ \rho + \mathbf{F}_{\rm visc}/\rho
\end{equation}
where $\BB$ is calculated through $\BB=\nabla\times\AAA$ and the current $\jj=\nabla\times\BB$.
The fluid velocity is $\vv$ and the convective derivative is denoted by $\tfrac{D}{D t} \equiv
\tfrac{\partial}{\partial t} + \vv \cdot \nabla$. The
simulation is isothermal, so the pressure is related to the density by $p=\rho c_s^2$ where
$c_s^2$ is the sound speed (set to unity).
The viscous force $\mathbf{F}_{\rm visc}$ is
calculated using the rate of strain tensor $\mathsfbi{S}$ whose indices are given by $S_{ij} =
\tfrac{1}{2}(\pder{u_i}{x_j} + \pder{u_j}{x_i}) - \tfrac{1}{3}\delta_{ij} \nabla \cdot \vv.$
through the equation:
\begin{equation}\label{eq:pencil_visc}
\mathbf{F}_{\rm visc} = \nabla \cdot 2 \nu \rho \mathsfbi{S},
\end{equation}

The continuity equation is implemented in terms of the logarithm of the density $\rho$ as
follows
\begin{equation}
\frac{D \ln \rho}{D t} = -\nabla \cdot \vv.
\label{eqn:continuity_eqn}
\end{equation}
Since the plasma is modeled as isothermal the equation of state does not need to be solved, so
the final equation is the induction equation, which in terms of the vector potential becomes
\begin{equation}
\frac{\partial \mathbf{A}}{\partial t} = \vv \times \mathbf{B} + \eta \nabla^2 \mathbf{A}
\label{eqn:induction_eqn}
\end{equation}
where we have chosen the Weyl gauge for $\AAA$ to simplify the equation and $\eta$ is the resistivity.

\refbold{These equations are solved on an Eulerian grid of $256^3$ grid points in a simulation box of
size 5. This puts the boundary at $R=2.5$. }
The simulation is run using perpendicular-field boundary conditions, by imposing a
vanishing parallel
component of the magnetic field on the boundary.
This boundary condition allows magnetic field to escape from the simulation volume.
The full field information is saved every 5 simulation time steps.
The simulation is initialized with a constant pressure $p=1$ throughout the volume and
the velocity field is zero. \refbold{The Alfv\'en speed $v_A$ in the initial field is
ranges from 0.28 to 1.1  (on axis) $B_0$ is varied between 0.05 and 0.2 and this makes it
the same order of magnitude as the sound speed $c_s^2=1$.}
We scale the time to a resistive timescale using $t_\eta = R^2_{\rm char}/\eta$.
For the characteristic length scale $R_{\rm char}$ we choose the distance of the
magnetic axis from the origin in the initial field, $R^*(0)=1$.
\refbold{The viscosity parameter $\nu$ is set to $2\times10^{-4}$ whereas the resistivity
    $\eta$ is varied from $2\times10^{-4}$ to almost an order of magnitude lower. This makes
the magnetic Prandtl number equal to unity or larger}

In our analysis of the simulation results we wish to extract the topological properties
of the magnetic field structures.
We do this by means of a Runge-Kutta field line integration.
From the resultant field line traces we can find the magnetic axis and the rotational
transform of the field line if it lies on a toroidal surface.
One implementation of this field line tracing is described in the supplemental material
of~\citep{smiet2015self}.
This method was used in Section~\ref{sec:resgrowth}-\ref{sec:analytics}.

Furthermore we have developed an alternative implementation in CUDA to run on graphics hardware.
The hardware accelerated trilinear interpolation and massive parallelization
allow for a high speedup compared to CPU-based field line tracing.
In the $(R,\!0,\!z)$-plane field lines are traced from a $1024\times 1024$ grid and
for every field line the rotational transform is calculated.

The evolution of the rotational transform profile can be seen in supplemental video 1.
This video shows the color-coded rotational transform, similar to the top row in figure
\ref{fig:cudanalysis}, for all times sequentially. Additionally, this video gives a good
indication of how the magnetic structure evolves in time.

The results of this analysis are shown in figure~\ref{fig:cudanalysis}, top row, for
the field with $\imath_0^*$=3, $B_0=0.05$ and $\eta=2\times 10^{-4}$.
Every single pixel in these images is the result of calculating the
rotational transform of a field line trace.
In this run the magnetic field remains axisymmetric and the magnetic axis
remains in the plane defined by $z=0$.
\refbold{This is the case for all runs presented in this paper, even though these
    symmetries are in no way enforced by the computational procedure. With higher values
    of $\imath_0^*$ the structure can become susceptible to a nonaxisymmetric kink
instability, which will be the subject of a future publication.}

Figure~\ref{fig:cudanalysis} (bottom) shows the rotational transform profile at different
times during the evolution of the configuration.
The rotational transform profile quickly shifts from positively curved to nearly flat with
a slight negative curvature. After this initial phase the rotational transform profile
remains nearly constant in space but decreases in time. In the next
sections we will study this decay.

In the rest of the paper we will analyze the change in rotational transform and the
location of the magnetic axis in time.
We will look at the effect of changing the resistivity $\eta$ and the initial
rotational transform $\imath_0^*$.

\section{Resistive growth and decay of rotational transform}\label{sec:resgrowth}

\begin{figure}\begin{center}
    \includegraphics[width=.6\columnwidth]{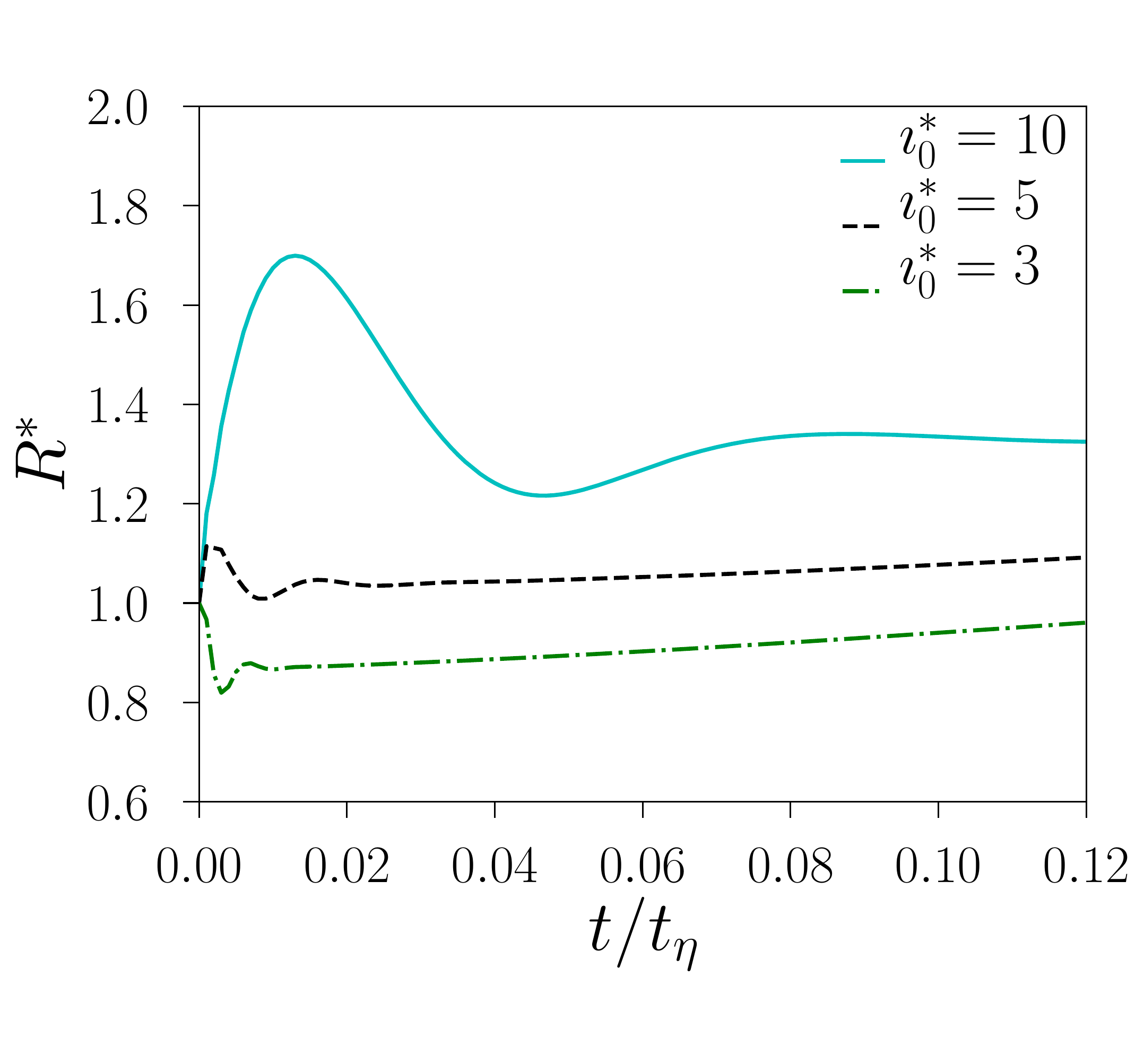}
    \caption{ Position of the
    magnetic axis in time for different values of $\imath_0^*$. The magnetic axis performs a
    damped oscillatory motion towards an equilibrium position which depends on
    the initial rotational transform, and then slowly grows.
    The parameters $B_0=0.05$ and $\eta=2\times 10^{-4}$ are fixed with $\imath_0^*$
    varied as shown.}\label{fig:radii}
\end{center}\end{figure}

In figure \ref{fig:cudanalysis} it is seen that the magnetic axis
first shifts inwards and then slowly moves back out.
We follow this dynamic of the magnetic structure by extracting $R^*$, the distance from the
origin to the magnetic axis as a function of resistive time.
This is done for three different values of $\imath_0^*$ at constant $\eta=4\times 10^{-4}$.
The results are shown in figure \ref{fig:radii}.
The structure relaxes to a radius which depends on $\imath_0^*$, and then slowly increases in size.

\refbold{As the structure grows, it has reached its equilibrium: the magnetic pressure
    pushes outwards, but the expansion is halted by the strong external pressure.
    A lowering of the pressure is observed in a toroidal region as we show here in
    figure~\ref{fig:flow}, and is described in more detail
in~\citep{smiet2015self,smiet2016ideal}. }

We can understand this initial relaxation qualitatively from
the interplay between magnetic tension and magnetic pressure.
Since the initial pressure is constant and the velocity is zero
the initial motion of the fluid is purely  due to the Lorentz
force $\jj\times\BB$.
A high value of $\imath_0^*$ results in a high poloidal field, \refbold{and magnetic
tension along the field lines squeezes the configuration into an expanding ring.}
The case of a low rotational transform will result in
stronger toroidal field, and the \refbold{magnetic tension will cause
the structure to contract}.
In figure \ref{fig:radii} we see that higher $\imath_0^*$ leads to an initial expansion, and
an equilibrium with a value of $R^*$ larger than $1$, whereas
low values of $\imath_0^*$ lead to an initial contraction.
$R^*$ performs a damped \refbold{Alfv\'enic} oscillation to the equilibrium position, and then slowly
grows.

\begin{figure}\begin{center}
    \includegraphics[width=.6\columnwidth]{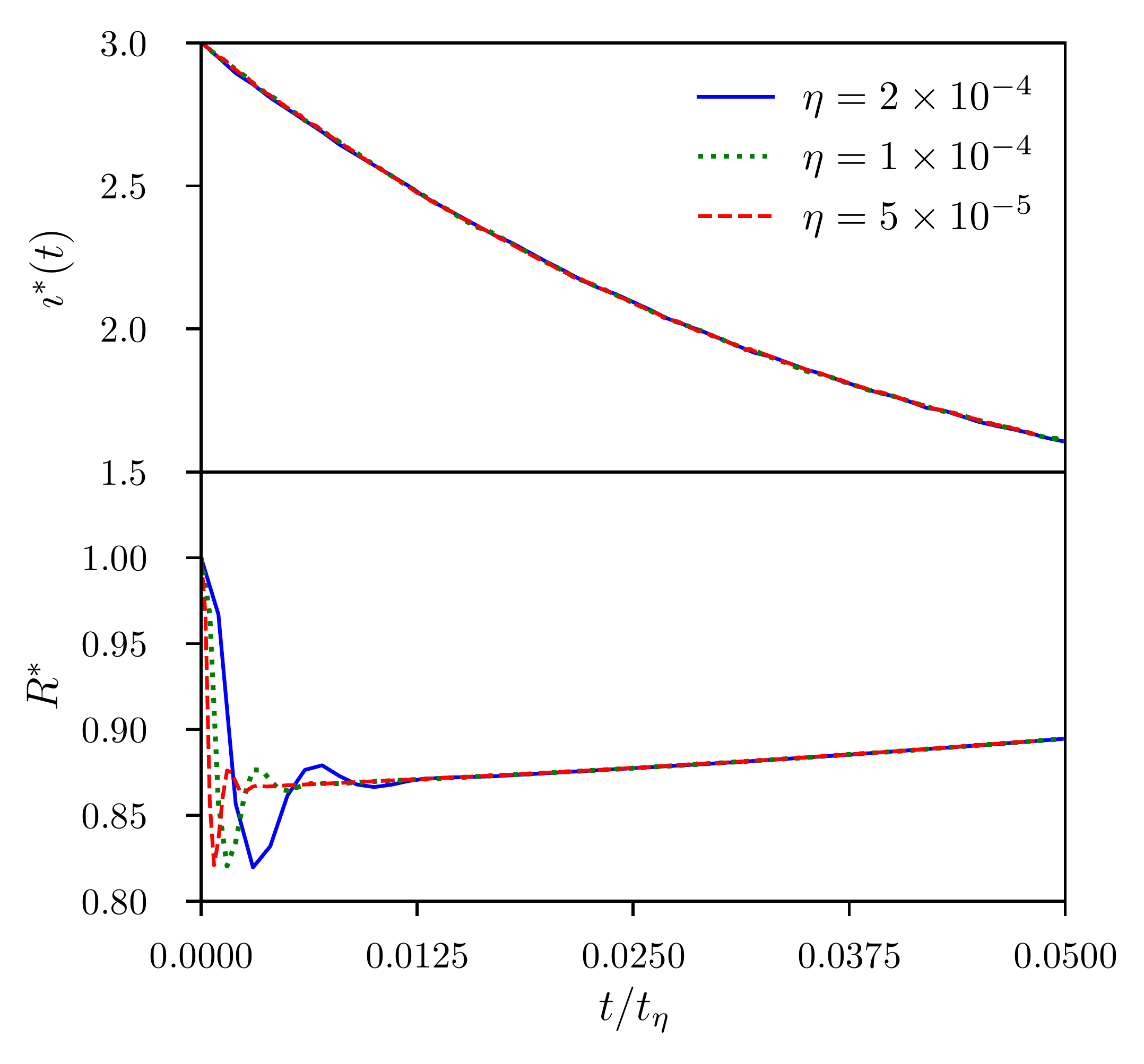}
    \caption{$\imath^*(t)$ and $R^*(t)$ as a function of resistive time for several different values of
    resistivity $\eta$. The initial rotational transform is set to $\imath_0^*=3$, and magnetic field strength
    $B_0=0.05$.
    The change of the rotational transform
    and the radius of the structure all behave identically on a resistive time scale.
    }
    \label{fig:compareplot}
\end{center}\end{figure}

The later evolution of the structure proceeds on a purely resistive time scale.
This is tested by simulating the evolution of the field with the parameters
$\imath_0^*=3$, $B_0=0.05$ and varying resistivity $\eta=2\times 10^{-4}, 1\times 10^{-4}$, and $5\times 10^{-5}$.
When rescaled to resistive time the growth of the structure and the change in rotational
transform all collapse to a single curve indicating a universal growth mechanism, as shown in figure
\ref{fig:compareplot}.

\begin{figure}
    \begin{center}
        \includegraphics[width=.6\columnwidth]{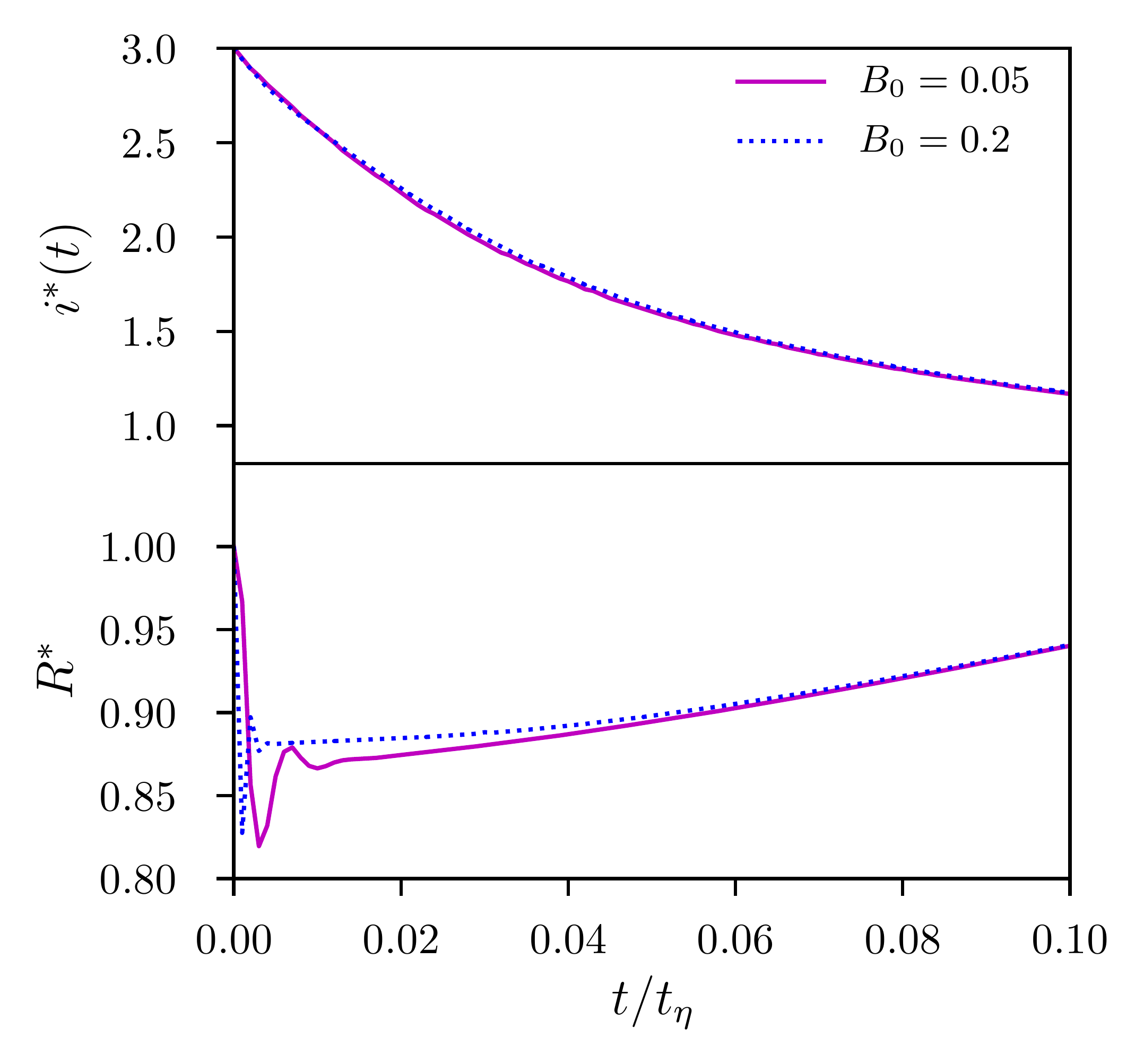}
        \caption{Magnetic decay of topologically identical structures with different
        initial magnetic field
        strength $B_0$. Despite the difference in magnetic field strength
        the change in rotational transform proceeds at exactly the same rate with identical
        equilibrium. Note that the initial oscillations towards the
        equilibrium radius occur on the Alfv\'enic timescale: the oscillations to the equilibrium
        configuration proceed at a four times faster rate when $B_0=0.2$ then when $B_0=0.05$.
        In these runs $\imath_0^*=3$ and $\eta=2\times10^{-4}$
        }
        \label{fig:compareplot_bmag}
    \end{center}
\end{figure}

The magnetic field strength $B_0$ does not
affect the equilibrium reached
or the rate of growth and change in rotational transform exhibited by these configurations.
This is shown in figure \ref{fig:compareplot_bmag}, where the field  with $B_0=0.05$ and $B_0=0.2$
are compared for $\imath_0^*=3$, $\eta=2\times10^{-4}$.
Despite a factor $4$ difference in the magnetic field strength the structures behave
identically except for the initial reconfiguration towards the equilibrium.
As this reconfiguration is mediated by magnetic forces, it proceeds on an Alfv\'enic
timescale linear in the magnetic field, $\tau_A=\frac{B}{\sqrt{\rho}}$.
It is therefore not surprising that the oscillation to the equilibrium $R^*$ lasts
about 4 times
longer for the field where the magnetic amplitude is a quarter of the strength.

\section{Pfirsch-Schl\"uter diffusion}\label{sec:analytics}
We can understand the structure growth and change in rotational transform through the effect of
finite resistivity on the plasma and the lowering of magnetic field strength through
resistivity.
In a perfectly conducting plasma a magnetic field is effectively `frozen-in' and moves with
the fluid motion~\citep{alfven1943existence, BatchelorFrozeIn1950RSPSA,
PriestReconnection2000}, thus there can be no net flow of fluid perpendicular to the field
lines if the magnetic configuration is static.
When resistivity is included this restriction is lifted and the fluid can slip against the
static magnetic field lines.
Field line slip is observed in many different scenarios and is one of the driving
mechanisms behind $2D$ reconnection~\citep{kulsrud2011intuitive}.
In the toroidal geometry of an operating tokamak, field line slip gives rise to
slow diffusion out of the toroidal flux surfaces. This process is called Pfirsch-Schl\"uter
diffusion~\citep{wesson2011tokamaks}.
This Pfirsch-Schl\"uter flow is directed outwards, in the direction of the pressure
gradient.

\begin{figure}
    \begin{center}
        \includegraphics[width=.6\columnwidth]{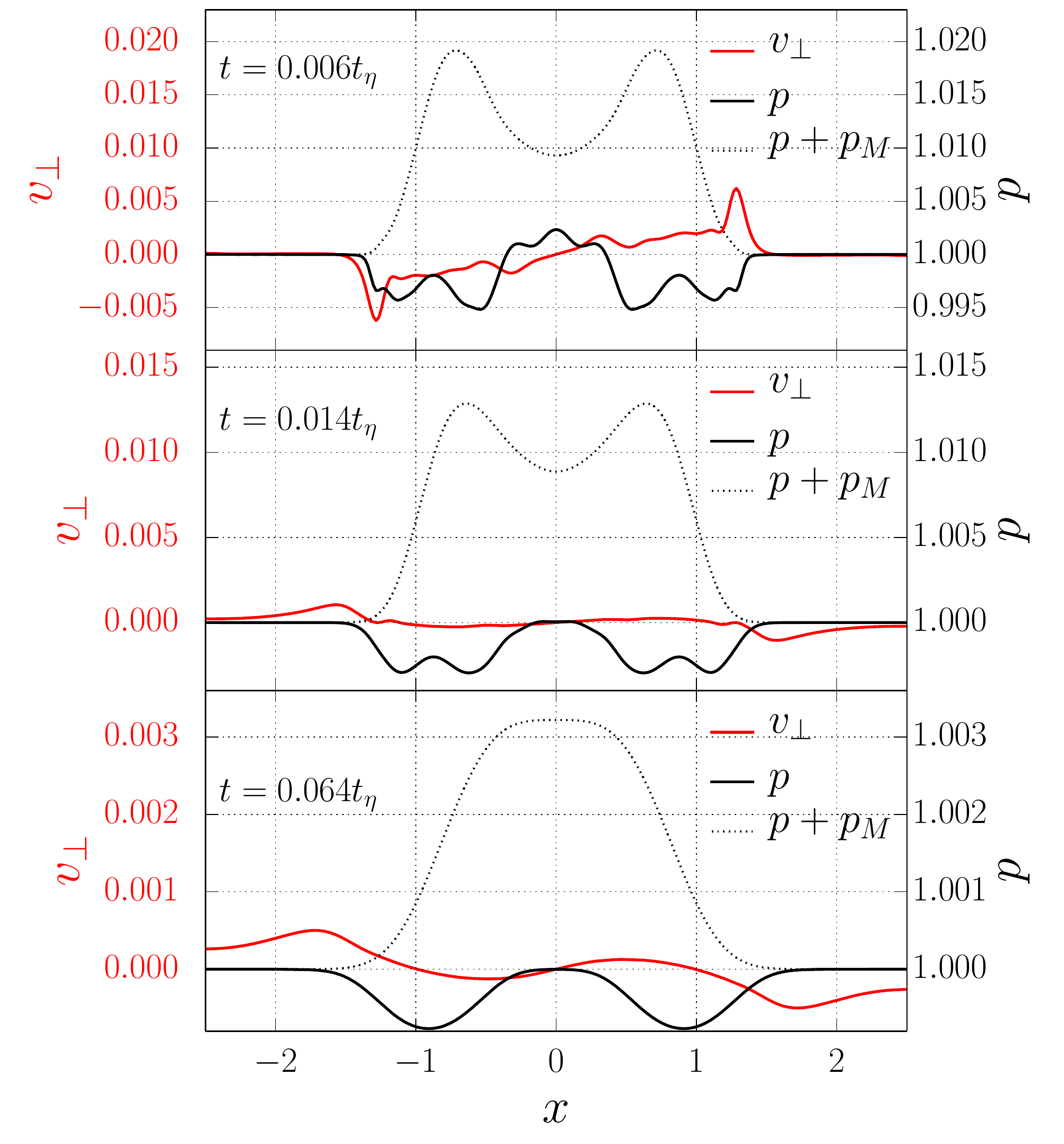}
        \caption{Fluid velocity $v_\perp$ (red) \refbold{pressure $p$ (black), and total
            pressure $p+p_M$} along the
            $x$-axis \refbold{at three different times corresponding to the second, third,
            and fifth top panel in figure \ref{fig:cudanalysis}.}
        The resistivity of this run was $\eta=2\times 10^{-4}$ and
        $\imath_0^*=3$.
        \refbold{During the initial reconfiguration, the pressure profile is irregular,
            but the total pressure is smooth. In the last panel the equilibrium has
        reached the state where the magnetic axis is}
        located at the minimum in pressure. The flow profile shows a net flow towards
        the magnetic axis \refbold{which is similar to Pfirsch-Schl\"uter diffusion}.
        }
        \label{fig:flow}
    \end{center}
\end{figure}

In the self-organized structures considered here a similar magnetic slip causes a diffusion of plasma fluid
into the magnetic structure.
This is shown in figure \ref{fig:flow}, where the flow field is plotted along the $x$-axis
together with the pressure profile.
The magnetic axis is located at the minimum of the pressure, and it is clearly seen how there is
net fluid flow directed towards the magnetic axis.
We suggest that the slight discrepancy between
the location of the magnetic axis and the zero of the velocity is due to the axis itself being
in motion.

\begin{figure}
    \begin{center}
        \includegraphics[width=.6\columnwidth]{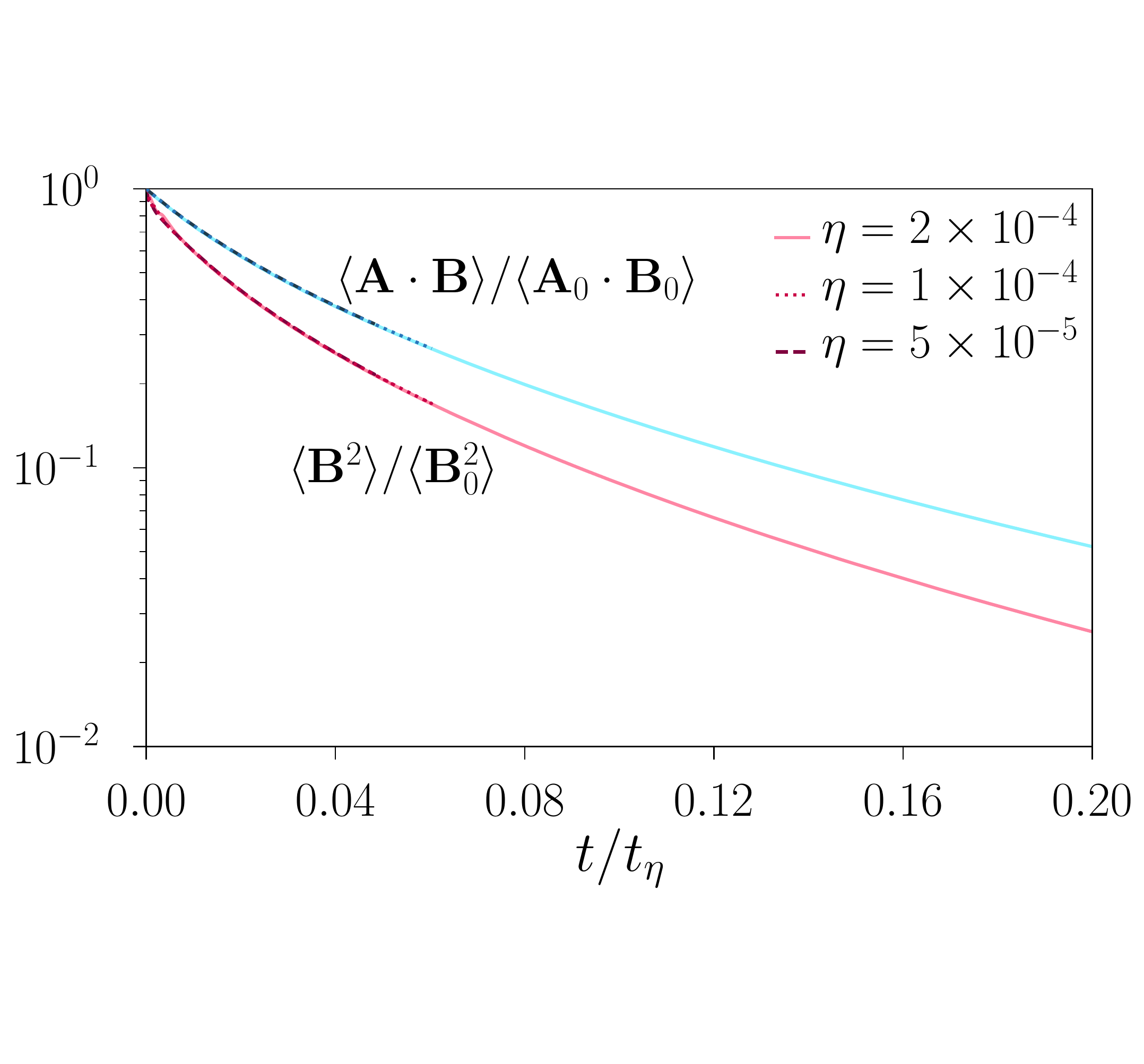}
        \caption{Decay of magnetic energy and decay of total helicity versus time for the runs with $\imath_0^*=3$ and $B_0=0.05$ for
        different values of resistivity $\eta$.
        }
        \label{fig:bmag5}
    \end{center}
\end{figure}

Whilst the fluid flow slowly penetrates the magnetic structure, the magnetic energy in the
structure is decreasing.
The decrease of total magnetic energy for the simulations with $\imath_0^*=3$ and $B_0=0.05$
is shown in figure~\ref{fig:bmag5}.

One important result to note is that the magnetic field strength decays fast compared to the
resistive decay time $t_\eta$, whereas in general the magnetic field strength is expected to
evolve as $\langle|\BB|\rangle\sim \langle B_0 \rangle e^{-t/t_\eta}$.
Here the magnetic energy has already decreased an order of magnitude
in only $0.1 t_\eta$. This is because the resistive losses are not the only mechanism through
which the magnetic field is lowered: During the evolution the entire configuration also
expands.
Even with zero resistivity such an expansion leads to a lowering of the magnetic field strength.
This can be seen as follows: since it is the flux through a co-moving surface that is
conserved, and if that surface expands, the
magnetic field strength lowers. This effect can also be seen in the zero resistance
simulations presented in~\citep{smiet2016ideal}.

\refbold{figure~\ref{fig:bmag5} also shows the evolution of magnetic helicity, which
    decreases at a slower rate than the magnetic energy. The slower decay of helicity can
    be seen as the result of the expansion, as the structure evolves with a self-similar
    shape. The magnetic energy is the integral of
    $(\nabla\times\AAA)\cdot(\nabla\times\AAA)$, whereas the integrand of the helicity
    integral, $\AAA\cdot(\nabla\times\AAA)$, involves one less spatial derivation. For a
    similar structure of a larger size, the magnetic helicity is thus larger.
    Note that we evolve these structures on a timescale larger than the timescales on
    which the helicity can be considered conserved, and that our initial condition is
    intentionally very regular. Therefore the localized reconnections which transform helicty
    between linking, writhe, and twist, which reconfigure the magnetic topology whilst
    leaving helicity mostly unchanged in turbulent Woltjer-Taylor type relaxation are
absent in these runs. See \citep{smiet2015self} for simulations where this equilibrium is
achieved through this more chaotic reconnection.}

\begin{figure}\begin{center}
    \includegraphics[width=.6\columnwidth]{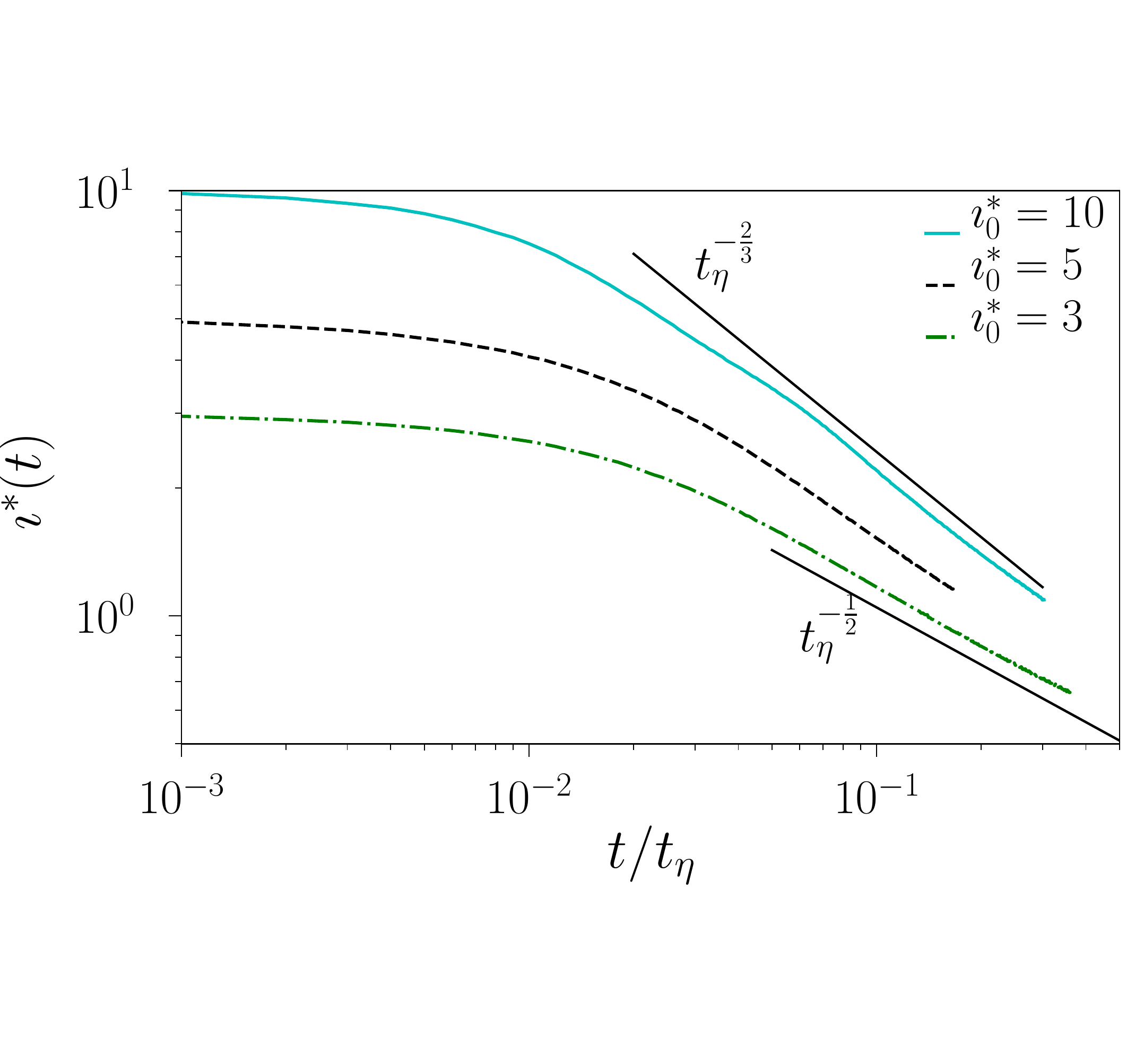}
    \caption{Time dependence of the rotational transform for different values of $\imath_0^*$.
    After the initial period where the magnetic structure re-organizes on an Alfv\'enic
    timescale, the rotational transform decays following a power law with characteristic
    exponent between $-2/3$ and $-1/2$.
    }
    \label{fig:decay}
\end{center}\end{figure}

Finally we look at the change of rotational transform in time, and how it depends on the
initial value for the rotational transform. The results are shown in  figure~\ref{fig:decay}.
From the asymptotic behaviour in the log-log plot we can see that the rotational transform
decays according to a power law instead of exponentially.
The characteristic exponent of this decay is different for runs with different
$\imath_0^*$.
The rotational transform between $t_\eta=0.05$ and $t_\eta = 0.1$ is fitted with a power law
$\imath^*(t)=a{t_{\eta}}^{-b}$ and a characteristic exponent of $b=0.664$ is found for the run with
$\imath_0^*=10$ and $b=0.48$ for the run with $\imath_0^*=3$.
Guides are drawn in figure~\ref{fig:decay} showing  $t_\eta^{-2/3}$ and $t_\eta^{-1/2}$ decay.

The lowering rotational transform is caused by the poloidal field decreasing faster than the
toroidal field. This is another indication that the lowering of field strength is primarily
caused by the
expansion of the structure caused by field line slip and not necessarily by resistive
decay.
\refbold{With resistive decay we indicate the exponential decay with characteristic decay
    time $t_\eta$ caused by resistivity on a static field configuration, solutions of
    equation~\eqref{eqn:induction_eqn} with $\vv=0$. Decay brought on by field line slip
    is caused by the non-zero Pfirsch-Schl\"uter flow, or conversely the motion of the
    field lines against the plasma. This causes the term $\vv$ to be non-zero and the term
    $\nabla^2\AAA$ to decrease (through expansion the gradients get smaller)
    in equation~\eqref{eqn:induction_eqn}, and thus
    the change in magnetic field strength can be fast compared to the resistive decay
    time.
}

As the structure expands the poloidal field strength is lowered by expansion in the
horizontal plane, as it is the poloidal flux passing through the circle defined by
the magnetic axis
which is conserved. The increase in area through which this flux passes is
proportional to $(R^* )^2$. The lowering of the toroidal field meanwhile is governed by the
change in area of the poloidal cross-section of the flux tube.
Expansion in this plane is constrained to the positive $R$-direction
and as such the area should go approximately linear in $R^*$. This constrained
expansion is attested by the D-shaped magnetic surfaces seen in the top row of
figure~\ref{fig:cudanalysis}.
Even though this explanation is
quite qualitative and does not take into account the shape of the surfaces and the
distribution of magnetic flux, it does explain why the rotational transform lowers according to
a power law. Furthermore, the difference between poloidal and toroidal expansion
correctly predicts a characteristic exponent of around $-1/2$.

It should be noted that the decay in rotational transform is fast
compared to the increase of the major radius of the structure; the rotational transform changes
by a factor of three in the time $R^*$ only changes a few percent.
This is important when considering this equilibrium as a model for magnetic clouds.

\section{Relation to magnetic clouds}

\refbold{The presented simulations of axisymmetric equilibria are more idealized than the
    situation encountered in the solar wind. As noted in the introduction, the plasma $\beta$
    in the solar wind is around 1, and a signature of a magnetic cloud is that the $\beta$
    drops below 1~\citep{zurbuchen2006situ}. In these simulations, the initial $\beta$ (taking the magnetic
    field strength and pressure on the axis) is around 25 when $B_0=0.05$ and 1.65 when
    $B_0=0.2$. However, as can be seen from figure~\ref{fig:compareplot_bmag}, when rescaled to
    resistive time the evolution of the fields is near identical. Simulations (not presented in
    this paper) have been performed at $\beta$ down to 0.7 which show the same evolution, but this
    is not a lower limit. The simulations show the set-up of a toroidal equilibrium against a
    background plasma with lower field strength, similar to the situation encountered in
    the solar wind.
}

\refbold{In the solar wind the Alfv\'en speed and the sound speed are the same order of magnitude
    with $v_s =37$ km $\rm{s}^{-1}$ and $v_A=47 $km $\rm{s}^{-1}$
    \citep{goedbloed2004principles}. Assuming a magnetic cloud size of approximately
    $10^{6}$ km (1 day passage for a probe travelling at 15 km $\rm{s}^{-1}$), the
    Alfv\'enic transit time is around 7 hours, so one dozen to several dozen Alfv\'en transit times
pass between coronal mass ejection and observation, a similar regime as is probed in the
simulation. The magnetic Prandtl number (ratio of kinematic viscosity to magnetic
diffusivity) for a hot thin plasma such as the interplanetary solar wind is
much higher than unity; $\sim 10^{14}$ (see \citet{brandenburg2005astrophysical}).
The plasma in the solar wind is thus in a regime here the viscous forces act faster than
the resistivity to allow for a similar self-organizing process as observed in the
simulations. One major difference is that $t_\eta$ for a magnetic cloud is
about $10^9$ years, much longer than the months for which a cloud can be observed before
it leaves the solar system. When the cloud is just ejected, $R_{\rm char}$ is smaller
and reconnection can occur (it must to trigger the ejection). Furthermore the more chaotic
solar wind allows for small-scale reconnection events which increase the effective
reconnection rate. Nevertheless, the extent to which the
rotational transform changes must be much smaller than the full evolution presented in
this paper.
}

There are several models for magnetic clouds discussed in literature, see for example
~\citep{burlaga1991magnetic} for an overview.
One of the more common approaches is to model a magnetic cloud as a long flux rope
extending from, and still
magnetically connected, to the surface of the sun.
Nevertheless there are several models that consider magnetic clouds as localized magnetic
excitations within the solar wind, with their magnetic field generated by internal currents.

Kumar and Rust describe a model for a magnetic cloud as an isolated, net current carrying
toroidal flux ring~\citep{kumar1996interplanetary}.
The magnetic field inside the ring is based on the force free Lundquist solution
valid for an infinite cylinder~\citep{lundquist1950magneto}.
As they themselves note, this cannot be an exact description, as the toroidal geometry
necessitates the existence of a hoop force such as described in~\citep{garren1994lorentz}.
In this model the plasma current is zero outside of the toroid, but the net current
through the toroid is non-zero such that it generates a force-free field in the
surrounding plasma.

There are several differences between Kumar and Rust's model and a magnetic cloud as a
self-organized structure we describe.
Firstly the magnetic field in their model is force free.
Such a field is not possible, as they note themselves because a current ring will always
experience a hoop force.
The rotational transform profile in their model is also very different.
In Lundquist solutions the axial and tangential fields (which, when the cylinder is translated to
a torus, correspond to the toroidal and poloidal directions respectively) are given by
Bessel functions. As such the rotational transform profile changes significantly from the
magnetic axis to the edge~\citep{bellan2000spheromaks}. The magnetic field outside the ring
is purely toroidal, which implies that the rotational transform goes to infinity.
Even though their model resembles the configuration we describe superficially, the rotational
transform profile is drastically different. Our simulations show that the profile quickly
flattens.

Another magnetic cloud model which resembles the configurations we observe is the
flare-generated spheromak model by Ivanov and Harshiladze~\citep{ivanov1985interplanetary} and
further explored by Vandas et al.~\citep{vandas1992magnetic}. They describe the clouds using the
spherical force free solution of Chandrasekhar and Kendall~\citep{chandrasekhar1957force}.
The magnetic topology in this solution also consists of field lines lying on nested toroidal
surfaces.
In this model the rotational transform profile of these force free solutions is also non-constant.

The resistive decay time for a structure with the characteristic length scale that is
reasonable for magnetic clouds, $R_{\rm
char}=10^6 {\rm km}$, is about $1.7\times10^9$ years~\citep{goedbloed2004principles}, so change in
the rotational transform due to resistive processes is expected to be small.
The process leading to the formation of the self-organized localized equilibrium
however takes place on a much faster timescale.
\refbold{Note that not just the Alfv\'enic oscillation towards equilibrium is fast; as seen
    in figure \ref{fig:cudanalysis} much of the change in rotational transform towards
equilibrium has already occurred at $0.006t_\eta$. This change, though fast, scales with
resistivity. If the resistivity is many orders of magnitude lower, as in the solar wind,
this change in
rotational transform can be expected to be much less rapid, and the cloud will still carry
much of
the topology it organized into when it was ejected.}
The evolution of the magnetic structure, after it is generated, can therefore be considered to be
approximately ideal, as is also assumed in the model by Ivanov and Harshiladze and the model by
Kumar and Rust.

Because the structure we describe does not rely on the assumption of force free fields, an
assumption that is not warranted in the $\beta\sim 1$ solar wind plasma, we
speculate that the magnetic structures described in this paper are a more realistic model
for localized magnetic clouds than the two others described above.

\section{Conclusions}
In this paper we have shown how a self-organizing equilibrium evolves on a resistive time
scale. In agreement
with our previous studies we find that the initially twisted flux tube reconfigures
on an Alfv\'enic timescale into an
axisymmetric Grad-Shafranov equilibrium characterized by a lowered pressure on the magnetic
axis.

In this paper we have described how the configuration evolves subsequently; the major radius
$R^*$ grows, and the rotational transform on the magnetic axis $\imath^*(t)$ lowers.
The rotational
transform profile, which initially had a high positive curvature, quickly evolves to a
almost flat and slightly negatively curved profile.

With the exception of the initial reconfiguration which proceeds on an Alfv\'enic timescale,
the evolution
is rather independent on the resistivity when scaled to a resistive timescale.
The growth of the structure can be understood as a Pfirsch-Schl\"uter-type slip of the field
lines against the plasma fluid background, or conversely the fluid slip against the field.

It is because of this growth of the structure that the magnetic field strength decays faster
than the resistive timescale. It is also this growth which allows the poloidal field to decay
faster than the toroidal field. We have also given a simple geometrical argument that explains why the
decay of the rotational transform behaves as a power law with characteristic exponent of the
order of $1/2$.

In this study we have limited ourselves to isothermal (constant resistivity) MHD evolution.
\refbold{The inclusion of temperature would result in a spatial variation of the Spitzer
    resistivity, which would quantitatively change the exact evolution, but the general
    aspects of the equilibrium and its evolution are underlied by geometrical principles
    and would remain unchanged. It is an interesting question whether the generation of a
flat rotational transform profile would remain robust under these conditions}

These results could help make predictions for the evolution of self-organized magnetic
equilibria in nature. In this paper we relate this structure to magnetic clouds. Other
applications for this model include AGN ejecta \citep{braithwaite2010magnetohydrodynamic}, and
one could possibly devise a scheme for pulsed nuclear fusion power generation
in which the plasma is confined in such a
magnetic structure, embedded in an extremely high fluid pressure environment.
When devising such a scheme it should be important to note that the decay time of the magnetic
field strength proceeds on a timescale much faster than the resistive decay time, and as such
that the `confinement' is a very transient phenomenon.

\section{Acknowledgements}
We wish to thank the anonymous referees for their careful review of our article.

This work is part of the Rubicon programme with project number 680-50-1532, which is (partly)
financed by the Netherlands Organization for Scientific Research (NWO).

DIFFER is part of the Netherlands Organisation for Scientific Research (NWO).

Notice: This manuscript is based upon work supported by the U.S. Department of Energy, Office
of Science, Office of Fusion Energy Sciences, and has been authored by Princeton University
under Contract Number DE-AC02-09CH11466 with the U.S. Department of Energy. The publisher, by
accepting the article for publication acknowledges, that the United States Government retains a
non-exclusive, paid-up, irrevocable, world-wide license to publish or reproduce the published
form of this manuscript, or allow others to do so, for United States Government purposes.

\bibliographystyle{jpp}
\bibliography{references}

\end{document}